\documentclass[sigconf, authorversion,nonacm]{acmart}

\AtBeginDocument{%
  }

\usepackage{amsmath,amsfonts}
\usepackage{algorithmic}
\usepackage{graphicx}
\usepackage{textcomp}
\usepackage{xcolor}
\usepackage{enumitem}

\usepackage{stfloats}

\usepackage{tikz}
\usetikzlibrary{shapes.geometric}

\usetikzlibrary{external}
\tikzsetexternalprefix{figs/}

\usepackage{pgffor}
\usepackage{pgfplots}
\usepackage{caption}  %

\usepackage{subcaption}
\usepackage{float}

\usepackage[capitalise]{cleveref}

\begin{document}

\title{$\mathbb{R}^4$: A \underline{R}acetrack \underline{R}egister File with \underline{R}untime Software \underline{R}econfiguration}

\author{Christian Hakert$^1$, Shuo-Han Chen$^2$, Kay Heider$^1$, Roland Kühn$^1$, Yunchih Chen$^1$, Jens Teubner$^{1,3}$, Jian-Jia Chen$^{1,3}$\\
$^1$: TU Dortmund University, Germany $^2$: National Yang Ming Chiao Tung University, Taiwan $^3$:LAMARR Institute for Machine Learning and Artificial Intelligence, Germany}

\renewcommand{\shortauthors}{Trovato et al.}

\begin{abstract}

    Arising disruptive memory technologies continuously make their way into the memory hierarchy at various levels. Racetrack memory is one promising candidate for future memory due to the overall low energy consumption, access latency and high endurance. However, the access dependent shift property of racetrack memory can make it easily a poor candidate, when the number of shifts is not properly reduced.
    Therefore, we explore how a register file can be constructed by using non-volatile racetrack memories with a properly reduced number of shifts.
    Our proposed architecture allows allocating registers in a horizontal or vertical allocation mode, where registers are either scattered across nanotracks or allocated along tracks. In this paper, we propose a dynamic approach, where the allocation can be altered at any access between horizontal and vertical. Control flow graph based static program analysis with simulation-based branch probabilities supplies crucially important recommendations for the dynamic allocation, which are applied at runtime. Experimental evaluation, including a custom gem5 simulation setup, reveals the need for this type of runtime reconfiguration.
    While the performance in terms of energy consumption, for instance, can be comparably high as SRAM when no runtime reconfiguration is done, the dynamic approach reduces it by up to $\approx 6\times$.
\end{abstract}


\maketitle

\pagestyle{plain}
\thispagestyle{plain}

\section{Introduction}
The rise of disruptive and emerging memory technologies makes them premier candidates to explore paradigm changing concepts of memory integration into modern computer architectures. Especially since extensive research on byte addressable non-volatile memories enabled them to be considered as mature candidates to serve as core memory components in a computer system, their potential of saving energy and unit cost makes them highly attractive \cite{pentecost:22}. Despite that, the actual non-volatility draws significant implications on system procedures related to persistency, which makes an exploration of possible integrations of NVMs across the memory hierarchy even more crucial.
A rather unique type of NVMs is formed by racetrack memories, which are characterized by their organization of storage domains into nanotracks and the requirement to \emph{shift} these nanotracks to an access head prior to an actual memory access \cite{PW2:2024}. Although these memories are highly attractive in terms of density, endurance, energy consumption and general latency, the shift induced overheads can become a significant problem in terms of additional energy and latency. Consequently, exploring the application of racetrack memories in a memory hierarchy crucially requires careful consideration and optimization of the induced overheads.

While racetrack memories have been explored widely across the memory hierarchy so far for background storage \cite{park2014accelerating}, main memory \cite{hu2016exploring} and caches \cite{wang2018ultra},
the application as CPU register has not been explored yet.
In this paper, to the best of our knowledge, we are the first to propose the realization of a \underline{r}untime \underline{r}econfigurable \emph{CPU \underline{r}egister file} with \underline{r}acetrack memories ($\mathbb{R}^4$). 
The nature of the racetrack organization into nanotracks with orthogonal access heads enables two allocation strategies for register contents, namely a horizontal and vertical allocation.  In the horizontal allocation, bits of a register are stored along nanotracks, while in the vertical allocation, single bits are scattered across nanotracks.

The vertical allocation favors frequent accesses to a small subset of registers, as one access port enables parallel access to multiple bits of a register without requiring additional shifts. Conversely, the horizontal allocation favors widely distributed register accesses across the register file, as it avoids long shifts prior to each access. 

This makes the choice between horizontal and vertical allocation an optimization problem, heavily dependent on the execution behavior of the running program. Since the execution behavior is highly dynamic, the allocation scheme should adapt to the current workload. 
To this end, our proposed $\mathbb{R}^4$ aims to provide architectural support for dynamic switching between vertical and horizontal allocations, leveraging the advantages of both allocation schemes.

To optimize this adaptive register allocation, we introduce a static analysis-based recommendation mechanism that guides the operating system to periodically reconfigure the runtime mapping of registers to nanotracks. This reconfiguration is done as a part of an interrupt handling routine, ensuring that the registers are restored to the updated mapping at the end of the interrupt without additional overhead for reformatting.
We create a \emph{control flow graph} on the compiled binary of a target application and apply a simulation of the induced overhead for both allocations. Combining all possible paths through the control flow graph for the subsequent time interval leads to a recommendation for the best expected allocation during this time interval. 
The recommendation is deployed together with the application binary to the target system, such that the system can load and apply the recommendations.

\noindent In short, this paper covers the following contributions:
\begin{itemize}[leftmargin=*]
    \item We propose a novel realization of a reconfigurable register file by racetrack memory, combining a horizontal and vertical allocation for application specific optimization (Section \ref{sec:assembly}).
    \item We provide characterization and analysis of the proposed register file in terms of induced shift overhead, energy consumption and execution latency by forming corresponding performance models (Section \ref{sec:costmodel}).
    \item We propose a novel static analysis control flow graph based recommendation system to perform optimized runtime allocation (Section \ref{sec:cfgopt}). The optimization improves performance by up to $\approx 6\times$.
    \item We conduct extensive evaluation of the proposed architecture on application benchmarks, featuring a custom gem5 simulator (Section \ref{sec:eval}).
\end{itemize}

\section{Related Work}
Exploration of racetrack memories, especially of domain-wall racetrack memories and of skyrmion racetrack memories have a far outreach in the related work. This exploration covers many levels across the memory hierarchy as of today. Background storage in the form of an SSD alternative is explored by Park et~al. for the application specific scenario of graph computation \cite{park2014accelerating}. A level above background storage, main memory is further explored to be replaced by racetrack memories. Hu et~al. propose main memory devices consisting of racetrack memories \cite{hu2016exploring}. Zhang et~al. further conduct parametrization and optimization of racetrack memory based storage devices \cite{zhang2018performance}.
Due to the potential low latency of racetrack memories, it is naturally considered as CPU-close memory in the form of caches and scratchpad memories. Wang et~al. propose a ring-shaped dense cache design \cite{wang2018ultra}. Hakert et~al. propose a specific racetrack scratchpad memory for the execution of decision trees \cite{hakert2021blowing,hakert2022rolled}.

To the best of our knowledge, the exploration of racetrack memories has not reached yet beyond the level of caches and scratchpad memories into CPU registers. Application of other non-volatile memory technologies for registers, however, received a certain focus in recent years.
The concept of a non-volatile register has traditionally been the cornerstone of the non-volatile processor (NVP), which operates in energy-harvesting devices that regularly run into power loss.  A non-volatile register is commonly implemented as a hybrid storage structure, utilizing volatile SRAM as the primary working memory and NVM as a secondary backup storage \cite{nvp:20}.  Li et al. \cite{NCFET:17} propose employing the inherent non-volatility of negative capacitance field-effect transistors (NCFETs). Thirumala et al. \cite{flipflop:18} propose a flip-flop register that operates in two modes, volatile and non-volatile, by utilizing a ferroelectric transistor and allowing dynamic switching.  Lee et al. \cite{leereram:17} introduce a non-volatile flip-flop based on ReRAM technology that has rapid power-off and rapid wake-up capabilities.

\section{System Model}
This paper conducts a basic exploration of the realization of a CPU register file by racetrack memory. Within this scope, this paper is neither limited to specific classes of CPUs, nor to specific realizations of racetrack memories.
In terms of the considered CPU architecture, we specifically target
CPUs with strict in-order register read and write.
This means, the transport of register contents to and from the pipeline must happen strictly in order. By this, we enfore a strict relation between the instruction stream and the register access sequence, allowing for analysis and optimization.
This, however, still allows a limited degree of out-of-order execution within the CPU pipeline.

Racetrack memory (RTM) is an emerging non-volatile memory technology with two main implementations: Domain Wall Racetrack Memory (DW-RTM) and Skyrmionic Racetrack Memory (SK-RTM) \cite{skm-write:24}. Both technologies store data as magnetic domains along nanoscale magnetic wires, known as \textit{nanotracks}.  Data is read/written by shifting the nanotracks through fixed access ports.
Memory devices for racetrack memories are usually realized as many nanotracks, which share aligned positions of access heads. This allows parallel access to the nanotracks. Shift pulses can either be synchronized in order to align all nanotracks at a unified offset or can be individual.
In this paper, we consider skyrmionic racetrack memories as the target system. SK-RTM features a unique property, where the stored information of a $1$ or $0$ bit is encoded by the presence or absence of skyrmions. Instead of destroying and regenerating skyrmions on a write access, this technology allows for a permutation write procedure, where existing skyrmions are shuffled at their storage domains and only complemented by newly generated skyrmions~\cite{PW1:2020,PW2:2024}. This allows us to consider a rather tight cost model in terms of energy consumption in this paper.

\section{Register File Architecture}
\label{sec:assembly}
\begin{figure}[]
    \centering
    \includegraphics[width=.48\textwidth]{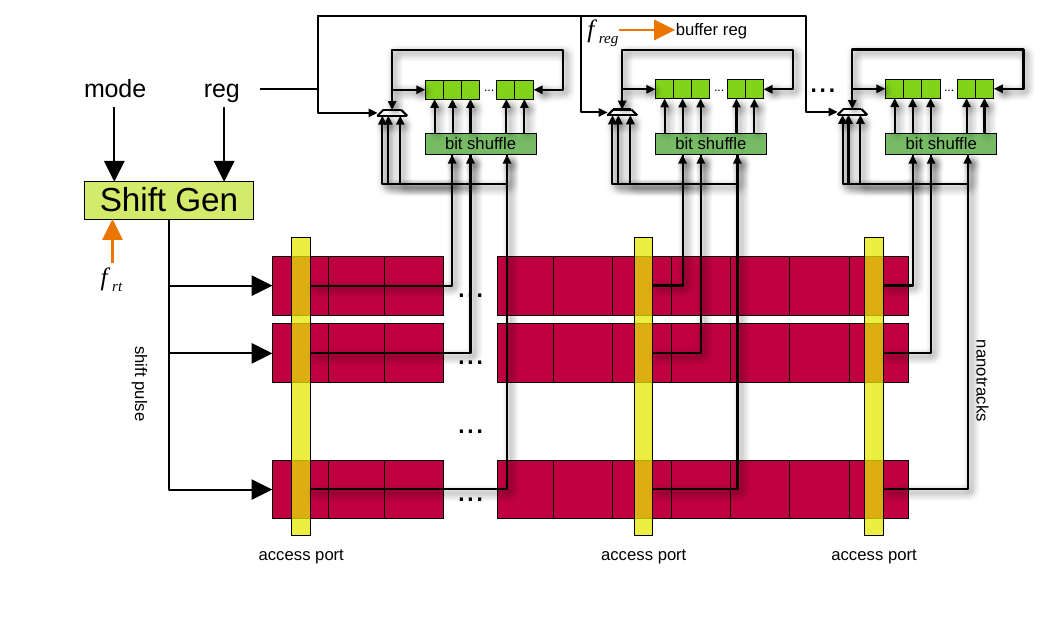}
    \caption{Overview of the Register File Architecture}
    \label{fig:regoverview}
\end{figure}
In order to assemble racetrack memory to a register file, two considerations have to be made: 1) the capacity requirement for the register has to be met and 2) the procedures of accessing single elements in the register file have to be designed and compared for the potential benefits and drawbacks. In this section, we present a generic architecture, which can be parametrized in order to derive certain access procedures and to make certain trade-offs.

\subsection{Assembly of the Register File}
\cref{fig:regoverview} illustrates an overview of the proposed assembly of racetrack memory into the register file. The basic memory component in this architecture is supplied by multiple racetrack nanotracks, illustrated in red.
All nanowires have the same length, i.e.~the same number of usable positions.
Denoting the number of nanotracks by $N$ and the amount of usable positions of each nanotrack by $W$, under the assumption that each position can store one bit, the total storage capacity is defined by $N\cdot W$. 
Although we do not assign fixed values for these numbers throughout this section, they are limited to an allocation, which meets the storage needs of the required register file. If, for instance, $R=32$ register with a width of $B=64$ bits are required, then $N\cdot W \geq 2048$. 
In addition, in order to simplify following consideration in the paper, we limit $N$ and $W$ to be powers of two (in \cref{sec:evaluation-nanotracks} we explore the effect of varying $N$ and $W$).

All nanotracks are equipped with $n_{ap}$ synchronous access ports, i.e.~the access ports are at the same position at every nanotrack. This number is limited to be a power of two as well. The access ports are placed equidistant at positions $\frac{W}{n_{ap}}\cdot(i+0.5) | i=0, ..., n_{ap}$, such that all access ports share the same number of surrounding positions. Each access port can access the positions of all nanotracks, that are currently aligned with the access port, i.e.~$N\cdot n_{ap}$ positions can be simultaneously accessed. 

For the horizontal allocation of registers into nanotracks,
at least one access port is placed per possible register, i.e. $n_{ap}\geq \lfloor\frac{W}{B}\rfloor$.
For the vertical allocation, similarly at least one access port is placed per possible register fragment (if one register spans multiple aligned positions) in the vertical allocation, i.e. $n_{ap}\geq \lfloor\frac{B}{N}\rfloor$.
It has to be noted at this point that the nanotracks physically need to contain more than the usable positions. Since each access port must be able to access $\frac{W}{n_{ap}}$ surrounding positions, the nanotrack may be shifted by an offset of up to $\frac{W}{n_{ap}\cdot 2}$ positions during operation. This requires additional overflow regions of $\frac{W}{n_{ap}\cdot 2}$ positions at each end of the nanotrack, which are not counted in the usable positions.

The access ports are connected to buffer register segments (illustrated in green) with two connection methods, dedicated to the vertical and horizontal allocation. For the vertical access method, all bits being accessed at an aligned position of all nanotracks are connected to the buffer through a shuffle circuit, that can reroute each bit to any position in the buffer segment. For the horizontal method, all bits from the access port are connected to a selection multiplexer/demultiplexer, which connects one of them to a serial connection of the buffer segment. The buffer segments further are able to perform a bidirectional roll operation through their serial access port. The number of buffer segments 
must meet the number of access ports.
The accumulated number of bits in all buffer segments is $B$ while the single segments have the same size. This allows to either transport all bits in the buffer register to or from a fixed aligned position of the nanotracks to the access ports when the bit shuffle is used. If the serial ports are used, this allows transport of all buffer bits to or from a minimum of $\max(1,\lfloor \frac{B}{W} \rfloor)$ nanotracks.

As a last component, we equip the register file with a shift generator, which is denoted by \emph{ShiftGen} in the figure. The shift generator generates a separate shift pulse for each nanotrack, such that each nanotrack can be individually shifted to an aligned position with the access ports. 

\subsection{Allocation and Access Modes}
The proposed architecture simultaneously allows serving for two orthogonal allocations of register contents to the nanotracks by setting the mode bit, shown in \cref{fig:regoverview}. We denote the allocations as vertical and horizontal. For the vertical allocation, nanotracks are always shifted synchronously, and the bits are transported through the bit shuffle circuits to and from the buffer segments. For the horizontal allocation, shifts are generated only for the nanotracks, that are chosen by the accessed register number.
Bits are transported through the serial access port and roll operations to and from the buffer segments.
In the vertical allocation, bits of a register can be accessed in parallel at an aligned position at the access ports, which favors repetitive accesses to the same register. In the horizontal allocation, all registers are aligned with at least one bit to at least one access port at any time, which favors distributed access to many registers.
In order to illustrate the two allocation and access modes in the following, we utilize a simple architecture in the following, which serves for $R=8$ registers with $B=8$ bits each. The register file is provided by $N=4$ nanotracks with a capacity of $W=16$ bits each. Each nanotrack is equipped with $n_{ap}=2$ access ports.

\subsubsection{Horizontal Allocation}
\newcommand{\emptyreg}{
    \node[draw,minimum width=8cm, minimum height=.5cm] (nt0) at (0,0) {};
    \foreach \n in {0,...,15}{\node[draw, minimum width=.5cm, minimum height=.5cm] () at (.5*\n-3.75,0) {};}
    \node[draw,minimum width=8cm, minimum height=.5cm] (nt1) at (0,-1) {};
    \foreach \n in {0,...,15}{\node[draw, minimum width=.5cm, minimum height=.5cm] () at (.5*\n-3.75,-1) {};}
    \node[draw,minimum width=8cm, minimum height=.5cm] (nt2) at (0,-2) {};
    \foreach \n in {0,...,15}{\node[draw, minimum width=.5cm, minimum height=.5cm] () at (.5*\n-3.75,-2) {};}
    \node[draw,minimum width=8cm, minimum height=.5cm] (nt3) at (0,-3) {};
    \foreach \n in {0,...,15}{\node[draw, minimum width=.5cm, minimum height=.5cm] () at (.5*\n-3.75,-3) {};}
    
    \node[draw,line width=2, color=yellow!80!black, minimum width=.5cm, minimum height=4cm] (ap0) at (-1.75,-1.5) {};
    \node[draw,line width=2, color=yellow!80!black, minimum width=.5cm, minimum height=4cm] (ap0) at (2.25,-1.5) {};
}
For the horizontal allocation, bits of a single register are allocated horizontally into positions of nanotracks. When the number of bits of the registers is larger than the width of the nanotracks, the registers span multiple nanotracks. These are consecutive and aligned at their access ports.
In case the bit width is lower, multiple registers are placed in a single nanotrack. Registers with consecutive numbers are placed in one nanotrack. The registers are distributed in such a way, that each register is aligned with the same number of access ports at any offset. When the number of access ports is greater than the number of registers per nanotrack, one register is accessed in parallel at different access ports and transported to or from separate buffer segments. When the number of access ports is equal to the number of registers, each register is only sequentially accessed through a single access port. By design, we enforce at least one access port per register in the horizontal allocation. The following illustration shows how the 8 registers are allocated horizontally.
\begin{center}
    \scalebox{.8}{
    \begin{tikzpicture}
        \emptyreg
        \node[minimum width=4cm, minimum height=.5cm,draw=red,fill=green!20!white, opacity=.6] () at (-2,0) {\scriptsize R0};
        \node[minimum width=4cm, minimum height=.5cm,draw=red,fill=green!30!white, opacity=.6] () at (2,0) {\scriptsize R1};

        \node[minimum width=4cm, minimum height=.5cm,draw=red,fill=green!40!white, opacity=.6] () at (-2,-1) {\scriptsize R2};
        \node[minimum width=4cm, minimum height=.5cm,draw=red,fill=green!50!white, opacity=.6] () at (2,-1) {\scriptsize R3};

        \node[minimum width=4cm, minimum height=.5cm,draw=red,fill=green!60!white, opacity=.6] () at (-2,-2) {\scriptsize R4};
        \node[minimum width=4cm, minimum height=.5cm,draw=red,fill=green!70!white, opacity=.6] () at (2,-2) {\scriptsize R5};

        \node[minimum width=4cm, minimum height=.5cm,draw=red,fill=green!80!white, opacity=.6] () at (-2,-3) {\scriptsize R6};
        \node[minimum width=4cm, minimum height=.5cm,draw=red,fill=green!90!white, opacity=.6] () at (2,-3) {\scriptsize R7};
    \end{tikzpicture}
    }
\end{center}
In order to perform an access to a specific register in this allocation, a shift pulse is generated for the nanotrack(s) containing the addressed register. In case the register spans multiple nanotracks, all containing nanotracks are included to the shift pulse synchronously. The nanotrack is first offset to the beginning of the addressed register content. Subsequently, the nanotrack is shifted position by position in order to make each bit of the register available at the access port. The bits are transported between the position and the serial port of the buffer segment. Roll operations in the buffer segment ensure a correct alignment before and after the access operation in the buffer segment. As a last step, the affected nanotracks are aligned back to their original offset by a corresponding shift pulse.

\subsubsection{Vertical Allocation}
In case of the vertical allocation, bits of a single register are distributed across as many nanotracks as possible in parallel with the access ports. In case the number of nanotracks is less than the number of bits in a register, the register is distributed to multiple positions in the nanotracks. By design, we enforce at least one access port per possible distributed register position. The distance of the positions is set to the same distance as the distance between access ports, such that one register can be accessed in parallel at the different access ports. In case there are more nanotracks than bits in a register, the corresponding bits are filtered out through the bit shuffle circuit.
For the example register file, each of the registers spans two positions in the nanotracks. Due to the existence of two access ports, one register can be entirely access in parallel.
\def\textstretch{.775cm}
\newcommand{\hreglab}[1]{\scriptsize #1\hspace*{\textstretch}#1\hspace*{\textstretch}#1\hspace*{\textstretch}#1}
\begin{center}
    \scalebox{.8}{
    \begin{tikzpicture}
        \node[minimum width=3.5cm, minimum height=.5cm,draw=red,fill=green!20!white, opacity=.6, rotate=90] () at (-3.75,-1.5) {\hreglab{R0}};
        \node[minimum width=3.5cm, minimum height=.5cm,draw=red,fill=green!20!white, opacity=.6, rotate=90] () at (0.25,-1.5) {\hreglab{R0}};
        \node[minimum width=3.5cm, minimum height=.5cm,draw=red,fill=green!30!white, opacity=.6, rotate=90] () at (-3.25,-1.5) {\hreglab{R1}};
        \node[minimum width=3.5cm, minimum height=.5cm,draw=red,fill=green!30!white, opacity=.6, rotate=90] () at (0.75,-1.5) {\hreglab{R1}};
        \node[minimum width=3.5cm, minimum height=.5cm,draw=red,fill=green!40!white, opacity=.6, rotate=90] () at (-2.75,-1.5) {\hreglab{R2}};
        \node[minimum width=3.5cm, minimum height=.5cm,draw=red,fill=green!40!white, opacity=.6, rotate=90] () at (1.25,-1.5) {\hreglab{R2}};
        \node[minimum width=3.5cm, minimum height=.5cm,draw=red,fill=green!50!white, opacity=.6, rotate=90] () at (-2.25,-1.5) {\hreglab{R3}};
        \node[minimum width=3.5cm, minimum height=.5cm,draw=red,fill=green!50!white, opacity=.6, rotate=90] () at (1.75,-1.5) {\hreglab{R3}};
        \node[minimum width=3.5cm, minimum height=.5cm,draw=red,fill=green!60!white, opacity=.6, rotate=90] () at (-1.75,-1.5) {\hreglab{R4}};
        \node[minimum width=3.5cm, minimum height=.5cm,draw=red,fill=green!60!white, opacity=.6, rotate=90] () at (2.25,-1.5) {\hreglab{R4}};
        \node[minimum width=3.5cm, minimum height=.5cm,draw=red,fill=green!70!white, opacity=.6, rotate=90] () at (-1.25,-1.5) {\hreglab{R5}};
        \node[minimum width=3.5cm, minimum height=.5cm,draw=red,fill=green!70!white, opacity=.6, rotate=90] () at (2.75,-1.5) {\hreglab{R5}};
        \node[minimum width=3.5cm, minimum height=.5cm,draw=red,fill=green!80!white, opacity=.6, rotate=90] () at (-0.75,-1.5) {\hreglab{R6}};
        \node[minimum width=3.5cm, minimum height=.5cm,draw=red,fill=green!80!white, opacity=.6, rotate=90] () at (3.25,-1.5) {\hreglab{R6}};
        \node[minimum width=3.5cm, minimum height=.5cm,draw=red,fill=green!90!white, opacity=.6, rotate=90] () at (-0.25,-1.5) {\hreglab{R7}};
        \node[minimum width=3.5cm, minimum height=.5cm,draw=red,fill=green!90!white, opacity=.6, rotate=90] () at (3.75,-1.5) {\hreglab{R7}};

        \node[minimum width=8.2cm, minimum height=.5cm,fill=white, opacity=1] () at (0,-0.5) {};
        \node[minimum width=8.2cm, minimum height=.5cm,fill=white, opacity=1] () at (0,-1.5) {};
        \node[minimum width=8.2cm, minimum height=.5cm,fill=white, opacity=1] () at (0,-2.5) {};
        \emptyreg
    \end{tikzpicture}
    }
\end{center}
In order to perform an access to a specific register in this allocation, all nanotracks have to be shifted synchronously to the offset of the addressed register first. The bits of the register are then transferred over the access ports and the bit shuffle circuit to or from the buffer segments. 
Roll operations are not required, since the bit shuffle circuits directly route the register bits to the correct positions. In addition, since all nanotracks are shifted synchronously, they don't have to be aligned again after the access. The global offset is stored in the shift generator.

\subsubsection{Switching Allocations}
In the proposed architecture, we assume the corresponding shift and roll operations, as well as the activation of the serial access port or the bit shuffle circuit all to be controlled by a single mode bit, which is one input of the register file. Since the architecture keeps all nanotracks aligned at one position after each access, no operations are required when the mode bit is changed to enforce a global alignment. The mode bit purely affects the procedure during an access, i.e. whether all nanotracks are shifted synchronously or a few nanotracks are shifted individually.
This means, that software can change the mode bit anytime, synchronized with the completion of the current access, without causing any overhead in the register file hardware. Naturally, the contents of the registers become invalid on a change of the mode bit, since other positions in the nanotracks correspond to a register after the mode switch. The software therefore has to assume all registers to become invalid on a change of the mode bit.
Consequently, system software has to perform an interrupt handler like procedure in order to change the mode bit, where old register contents are stored to main memory, the mode bit is changed, and the register contents are restored from main memory. Integrating the mode bit change into existing interrupt handlers by altering the mode bit just before the restore procedure makes this process more efficient.

\section{Cost Model}
\label{sec:costmodel}
In order to conduct further analysis and lay out the basis for optimization by runtime reconfiguration of the register file allocation, we formulate a cost model in this section. Generally, the register file can operate under two access modes, where switching between the access modes comes at no additional cost in the hardware. Hence, we formulate each a cost model for a sequence of register accesses under the vertical and under the horizontal allocation in the following. 

\subsection{Shift Cost Model}
Due to the inherent nature of racetrack memories, the induced shifts always depend on the previously aligned position of the nanotracks. The hardware has to generate a shift pulse, that shifts the nanotracks exactly by the required offset to align them with the target position. Consequently, in order to build a shift cost model, the input to the model is considered to be the last accessed register $reg_{old}$ and the next to be accessed register $reg$. The shift cost model then delivers the required amount of shifts for the indicated access. By summing up the results of the model over a sequence of accessed registers, the total shifting cost can be determined.

\subsubsection{Horizontal Allocation}
Considering the required shifts for accesses in the horizontal allocation, two cases must be distinguished. 1) one or multiple registers are placed within a single nanotrack, i.e., $B\leq W$. 2) one register is stored across multiple nanotracks, i.e., $B>W$.

\textit{Case 1: $B\leq W$:} when at least one register is placed entirely in one nanotrack, only the nanotrack corresponding to the register needs to be shifted. Additionally, each nanotrack has at least as many access ports as registers allocated in the horizontal allocation. This ensures that at least one access port is always aligned at an offset within the target register. Therefore, the access operation requires shifting to the beginning and end of the register and then returning to the original position. If multiple access ports are aligned within the width of a single register, the number of shifts is shared equally in parallel among the access ports. In greater detail, $\frac{W}{B}$ register are placed in one nanotrack, where $\frac{n_{ap}\cdot B}{W}$ access ports are dedicated to each register in the nanotrack. 
It should be noted, since all dimensions of the register file architecture are enforced to be powers of two, the fractions are either less than one or integer numbers.
The $B$ bits of each register are shared by this number of access ports. This results in the following required amount of shifts:

\begin{equation}
    \begin{aligned}
        B\leq W \Rightarrow \mathbb{S}_h(reg_{old},reg)=\left(\frac{B\cdot W}{n_{ap}\cdot B}-1\right)\cdot 2\\
        \Leftrightarrow\\
        B\leq W \Rightarrow \mathbb{S}_h(reg_{old},reg)=\left(\frac{W}{n_{ap}}-1\right)\cdot 2
    \end{aligned}
\end{equation}
It can be observed that this allocation neither is dependent on the previously accessed register, nor on the currently accessed register. The shifting cost therefore is independent of the register access sequence.

\textit{Case 2: $B>W$}: when one register spans across multiple nanotracks, the shifts are not only equally distributed between access ports, but also between the $\frac{B}{W}$ nanotracks, allocated by the register. However, also all of the $\frac{B}{W}$ registers have to be shifted to perform the access. The shifts along all $W$ bits within each nanotrack are shared by all available access ports.
\begin{equation}
    \begin{aligned}
        B > W \Rightarrow \mathbb{S}_h(reg_{old},reg)=\left(\frac{W}{n_{ap}}-1\right)\cdot 2\cdot \frac{B}{W}
    \end{aligned}
\end{equation}
Both cases can be unified in one equation:
\begin{equation}
    \begin{aligned}
        \mathbb{S}_h(reg_{old},reg)=\left(\frac{W}{n_{ap}}-1\right)\cdot 2\cdot \max \left(1,\frac{B}{W}\right)
    \end{aligned}
\end{equation}

\subsubsection{Vertical Allocation}
The case of vertical allocation can be analyzed similarly to the horizontal allocation.
 
\textit{Case 1: $B< N$:} In this case, multiple registers can be stored at a single aligned position across all nanotracks, i.e., $N > B$, $\frac{N}{B}$ registers can be stored at a single position. Consequently, the distance between the positions of two registers is scaled by a factor of $\frac{B}{N}$.
Depending on the number of access ports distributed across the nanotracks, $n_{ap}\cdot\frac{N}{B}$ registers can be accessed at a single aligned position across all access ports.
Since all nanotracks are shifted synchronously, the shifting cost is therefore given by:

\begin{equation}
    \begin{aligned}
        N>B\Rightarrow \mathbb{S}_v(reg_{old},reg)=
        \\|\lfloor reg\cdot \frac{B}{N\cdot n_{ap}}\rfloor-\lfloor reg_{old}\cdot\frac{B}{N\cdot n_{ap}}\rfloor|\cdot N
    \end{aligned}
\end{equation}

\textit{Case 2: $B\geq N$}:
In this case, an aligned position across all nanotracks can hold at most one register, i.e. $N\leq B$,
a register is distributed across $\lceil\frac{B}{N}\rceil$ positions in the nanotrack. There must be at least one access port per position, i.e. $n_{ap}\geq \lceil\frac{B}{N}\rceil$. Therefore, the shifting cost is given by shifting all nanotracks from the previous offset to the current offset within at most $R$ steps. If there are more access ports than $\frac{B}{N}$, the aligned positions are shared across these access ports, reducing the shift distance with a factor of $\frac{B}{N\cdot n_{ap}}$:
\begin{equation}
    \begin{aligned}
        N\leq B\Rightarrow \mathbb{S}_v(reg_{old},reg)=
        \\|\lfloor reg\cdot \frac{B}{N\cdot n_{ap}}\rfloor-\lfloor reg_{old}\cdot\frac{B}{N\cdot n_{ap}}\rfloor|\cdot N
    \end{aligned}
\end{equation}
The shifting cost remains the same for both scenarios. In the vertical allocation, the sequence of register accesses can significantly impact the overall shifting cost. Consecutive accesses to the same register, on the other hand, incur no additional shifting cost.

\subsection{Energy Model}
Although the previously introduced concepts in this paper are not specific to a realization of racetrack memory, we narrow down the subsequent explorations of this paper to a specific realization in order to construct realistic energy and latency models. We consider skyrmion racetrack memory in the following.
As a key difference, skyrmion racetrack memory has the possibility to not only apply a central shift pulse for one nanotrack, but also apply a shift pulse at each access port for a subsegment of the nanotrack. Although this capability has no impact on the positioning shifts, as introduced before, it impacts the efficiency of read and write procedures, which is discussed in the following.

The energy model can be derived based on the number of required shift, detect, insert and, remove operations, where each operation is associated with a per access port and per position energy consumption, namely $E_d,E_s,E_r,E_i$. Notably, these four operations are considered as the basic operations of racetrack memories and assumed to have near-constant energy consumption (see \cref{skyr_energy_opera}).
For the horizontal layout, this paper considers a permutation write strategy to update the content of a register, as the permutation write strategy delivers the best energy efficiency by repermutating existing data bits to compose new data patterns \cite{PW1:2020}. 
For the vertical layout, the na\"ive write strategy is utilized to remove all existing data bits and insert data bits at corresponding offsets of new data patterns. On the other hand, reading registers only requires shifting data bits across access ports and does not induce insert or remove operations.

\begin{table}[htb]
	\centering
	\begin{tabular}{l|cccc}
		Operations & $E_{d}$ & $E_{s}$ & $E_{r}$ & $E_{i}$ \\
		\hline
		Energy & 2~fJ & 20~fJ & 20~fJ & 200~fJ \\
	\end{tabular}
	\caption{Energy consumption of SK-RM operations}
	\vspace*{-1cm}
	\label{skyr_energy_opera}
\end{table}

\subsubsection{Horizontal Allocation}
In addition to the required shifts to align all bits of a register with an access port,
the bit difference when writing a new register value is the key when driving the energy model for the horizontal layout with the permutation write strategy. This is because excessive data bits need to be removed, while additional data bits should be inserted only when there are insufficient data bits from the original data pattern. To determine a 1 or 0 bit in the original register value, $B$ detect operations are required. 
Assuming the number of 1 bits in the original and to-be-updated data patterns are $Q$ and $Q^\prime$, if there are excessive data bits, the number of excessive data bits, $j$, can be calculated as $max(Q-Q^\prime, 0)$, which should be removed with one remove and one additional shift operations. On the other hand, when inserting data bits is needed, the number of to-be-inserted data bits, $k$, can be formulated as $max(Q^\prime-Q, 0)$. Notably, since permutation writes and a register on multiple tracks cannot share a unified shift pulse, individual shift pulses have to be applied at each access port.
This requires separate shift pulses for all access ports, being aligned with one register. Other access ports in the affected nanotracks, however, do not cause additional shift energy consumption, since the shift pulse can transit through them.
Therefore, the shifting cost of $\mathbb{S}_v(reg_{old},reg)$ is multiplied by $n_{ap} \cdot min(1,\frac{B}{W})$ for computing the energy consumption of a write operation in the horizontal allocation. Considering the above procedure and denoting the energy consumption of racetrack memory operation as $E_{s}$, $E_{d}$, $E_{i}$, and $E_{r}$, the energy model of horizontal model with permutation writes can be formulated as follows.

\begin{equation}
\mathbb{E}^{w}_{h} = \mathbb{S}_h(reg_{old},reg) \cdot n_{ap} \cdot min(1,\frac{B}{W}) \cdot E_{s} + B \cdot E_{d} + k \cdot E_{i} + j \cdot (E_{s}+E_{r}) 
\label{eq_enhw}
\end{equation}

To provide more intuition for the permutation write \cite{PW1:2020}, especially the term $j \cdot (E_{s}+E_{r})$, consider the following example:  Given, a single nanotrack with one access port and 8 positions, encoding two 4-bit wide registers with the content
\\
\texttt{{\hphantom{1111}1100\fbox{0}101}},
\\
where the access port is indicated to be aligned with the fifth position. The goal is to update the second register to achieve a value of 
\\
\texttt{{\hphantom{1111}1100\fbox{0}010}}. 
\\
The required shifts to achieve this alignment are included as one component in the first part of \cref{eq_enhw}.
The shifting cost for the horizontal allocation includes shifting from one end of a register to the other and back. To illustrate, the register is shifted to the end, resulting in 
\\
\texttt{{\hphantom{1}1100010\fbox{1}}}.
\\
Then, a zero bit is inserted at the access port, yielding 
\\
\texttt{{\hphantom{1}1100010\fbox{1}0}}.
\\
The next 1 bit is retained, resulting in 
\\
\texttt{{\hphantom{11}110001\fbox{0}10}}.
\\
The next bit is also kept, leading to 
\\
\texttt{{\hphantom{111}11000\fbox{1}010}}.
\\
Finally, the last 1 bit is obsolete and needs to be removed while shifting back to the original alignment, resulting in 
\\
\texttt{{\hphantom{1111}1100\fbox{0}010}}.

This example demonstrates that due to the obsolete 1 bit, an additional shift was required to establish the new 0 bit, alongside the energy overhead of actually removing the skyrmion.
For reading a register value in the horizontal allocation, no permutation write needs to be applied. The energy consumption then only consists of the required shifts, which can be performed in parallel for the entire nanotrack, and the detection of the single bits at the access ports.

\begin{equation}
\mathbb{E}^{r}_{h} = \mathbb{S}_h(reg_{old},reg) \cdot E_{s} + B \cdot E_{d}
\end{equation}

\subsubsection{Vertical Allocation}
Considering the na\"ive write strategy for vertical allocation, the write energy is primarily dominated by the number of bits, $B$, in each register. When updating a register in the vertical allocation, the target register $\text{reg}$ is first aligned with the corresponding access ports, causing the previously described shifts. Then, $B$ remove operations are applied to delete the original data, followed by up to $B$ insert operations to write the updated data. Assuming the number of 1-bits in the new data pattern is $Q^\prime$, the energy consumption for insertion can be calculated accordingly. 
Similarly, reading a register in vertical allocation involves aligning the register to an access port, followed by $B$ detect operations.
In summary, the energy consumption for write and read operations in vertical allocation can be expressed as:

\begin{equation}
\mathbb{E}^{w}_{v} = \mathbb{S}_v(reg_{old},reg) \cdot E_{s} + B \cdot E_{r} + Q^\prime \cdot E_{i}
\end{equation}

\begin{equation}
\mathbb{E}^{r}_{v} = \mathbb{S}_v(reg_{old},reg) \cdot E_{s} + B \cdot E_{d}
\end{equation}

\subsection{Latency Model}
Following the same set of operations mentioned above, the latency model for the horizontal and the vertical allocation can also be derived accordingly. 
The latency model again is based on an induced latency per detect, shift, remove and insert per access port and position. The corresponding latency components, namely $L_d,L_s,L_r,L_i$ are given in \Cref{skyr_laten_opera}.
Namely, for the horizontal and the vertical allocation, the latency model of the permutation write and the na\"ive write strategy are given in this section. Notably, the main difference is that, while the energy consumption of each operation is counted cumulatively, the latency of skyrmion operations is counted only once if operations are conducted in parallel.

\subsubsection{Horizontal Allocation}
To update registers through the permutation write strategy, positions in the nanotrack are processed separately. Although multiple access ports may be aligned with a single register, the permutation write causes local shifts at the access port, which blocks other access ports from local operation during this time. Therefore, the permutation write operation cannot be parallelized across multiple access ports.
First, the to-be-updated positions are shifted out and detected by $B$ detect operations. Then, while previous skyrmions are shifted back for composing the new data pattern, there could be remove or inject operations if there are excessive or insufficient skyrmions. Accordingly, the latency model for the register write/update is as follows, while $L_{s}$, $L_{d}$, $L_{i}$, and $L_{r}$ represent the latency of a shift, detect, insert, and remove. At any given time, only one of $j$ or $k$ has a value greater than 0, as excessive skyrmions and insufficient skyrmions are mutually exclusive conditions. Therefore, the latency is not over-estimated.

\begin{equation}
\mathbb{L}^{w}_{h} = \mathbb{S}_h(reg_{old},reg) \cdot L_{s} + B \cdot L_{d} + k \cdot L_{i} + j \cdot (L_{s}+L_{r}) 
\end{equation}

On the other hand, reading a register is also carried out following the similar approach of shifting a register through the access ports with only detect operations. The model can be summarized as:

\begin{equation}
\mathbb{L}^{r}_{h} = \mathbb{S}_h(reg_{old},reg) \cdot L_{s} + B \cdot L_{d}
\end{equation}

\subsubsection{Vertical Allocation}
For the vertical allocation, skyrmion operations to a vertically stored register are conducted in parallel. That is, for instance, all nanotracks are shifted at the same time and all skyrmions on the access port can be removed and injected in parallel. Therefore, the latency of skyrmion operations is counted only once while there are actual multiple operation conducted in parallel. The write/update, $\mathbb{L}^{w}_{v}$, and read latency, $\mathbb{L}^{r}_{v}$, can be summarized as follows.

\begin{equation}
\mathbb{L}^{w}_{v} = \mathbb{S}_v(reg_{old},reg) \cdot L_{s} + L_{r} + L_{i}
\end{equation}

\begin{equation}
\mathbb{L}^{r}_{v} = \mathbb{S}_v(reg_{old},reg) \cdot L_{s} + L_{d}
\end{equation}

\begin{table}[htb]
	\centering
	\begin{tabular}{l|cccc}
		Operations & $L_{d}$ & $L_{s}$ & $L_{r}$ & $L_{i}$ \\
		\hline
		Latency & 0.1~ns & 0.5~ns & 0.8~ns & 1~ns \\
	\end{tabular}
	\caption{Latency SK-RM operations}
	\label{skyr_laten_opera}
\end{table}

\section{Optimization by CFG Analysis}
\label{sec:cfgopt}
Given the previously described cost models, optimization for the runtime reconfiguration can be conducted.

\subsection{Software Remapping Architecture}

The mode control input for the register file, determining whether register are accessed in the horizontal or vertical allocation has to be kept constant until system software actively triggers a change and handles the corresponding data migration procedures.
Therefore, we propose using a central system configuration register to supply the allocation mode bit, ensuring that the access mode remains unchanged until software explicitly modifies this bit in the system control register. Additionally, we propose a software setup,
in which this control bit is modified within an interrupt handler, exploiting the register store and restore procedures of interrupt handling to safe a majority of the store and restore overheads.

Reconfiguring the register allocation during interrupt handling also enables a straightforward optimization approach, where an optimal allocation is derived for each interval between two interrupts.
With a predefined interval based on a fixed number of executed instructions, we can perform offline analysis and optimize the allocation according to the current position in the application and the length of the next interval. These offline optimization results, represented as recommendation bits, are deployed with the application, with each instruction carrying a single recommendation bit.

We propose a hardware peripheral that is triggered after each fixed number of instructions and loads the recommendation bit corresponding to the current program counter. This peripheral checks if the recommendation bit matches the currently active allocation or if an allocation change is required. In the latter case, the hardware peripheral triggers an interrupt to alter the allocation.
The interrupt handler then only needs to flip the current allocation bit, without loading additional recommendation bits. During interrupt handling, the storing and restoring of registers automatically reformats them to align with the new allocation.

The space overhead for recommendation bits is limited to \(\frac{1}{32}\) of the text segment in a 32-bit instruction set architecture. The recommendation bits can be loaded into a dedicated read-only memory upon program loading, accessed solely by the hardware peripheral.
If the application’s text segment becomes large, and the recommendation bit storage grows correspondingly, a caching scheme similar to a TLB could be integrated into the hardware peripheral to manage recommendation bit access efficiently.

\subsection{Static Binary Analysis}
In order to derive the offline optimized recommendation bits, which are deployed together with the target application to the system, we present a probability annotated static analysis approach in this subsection, which builds the control flow graph (CFG) on the compiled binary of the target application and derives expected optimal register file allocations.

When deriving an offline optimized recommendation, the main focus is put to the sequence of register accesses to be expected at a certain position in the application for the next interval. Since our target system features strict in order register accesses, the sequence of register accesses can be directly inferred from the instruction stream. For this, it is assumed that source registers are loaded in an earlier pipeline stage than destination registers are written. Consequently, considering the set of source and destination registers of each instruction can reveal the sequence of register accesses for this particular instruction. In order to gather this information, we utilize a disassembler to transform the application binary into a parsable format. Binary lifters, such as BAP \cite{brumley2011bap}, may be desirable for this process due to their feature richness, however, they do not reveal the adequate degree of detailed information, since they lift the binary program into an intermediate language, which does not necessarily maintain the name and sequence of register accesses. Consequently, we only utilize the disassembler from the GNU debugger (gdb) for the proposed analysis.

With the help of the gdb disassembler, we iterate over the application binary instruction by instruction. For each instruction, we separate the used register into a set for source registers and a set for destination registers. We annotate each single instruction with the corresponding sets. Next, we group the instructions to basic blocks. Branch instructions with a static branch target within the application binary are directly interpreted by gdb, which allows us to annotate these branches in a control flow graph. We further detect linked branches by the corresponding mnemonics and push the potential link targets through the control flow graph until a return instruction is detected. 
In order to get an estimate of the probability of a branch being taken or not taken, we simulate the first $1\,000\,000$ instructions of the application and count the empirical probabilities of conditional branches being taken. We annotate the CFG further with these probabilities.

Although this control flow graph cannot guarantee full coverage, it reveals the control flow along with the set of source and destination registers for each instruction based on a given application binary, largely independent of the compiler used. Since this control flow graph is intended for optimization purposes rather than functional correctness, we consider the risk of incomplete coverage acceptable.

\subsection{Derivation of Recommendation Bits}
Based on the derived control flow graph, we consider each single instruction as a starting point and derive a recommendation for each instruction to be deployed with the application. For this, we consider the length of each interval as a number of instructions and span all possible paths through the CFG with the desired target length starting at the position of the current instruction. For each path, we apply a simulation of the shift cost model as the objective function for both register file allocations based on the source and destination register set and derive two score values per path for the horizontal and vertical allocation, respectively. The lower scored allocation is considered to be the optimal choice for each path. Each path is weighted with the absolute empirical branch probability for this path to be taken. The overall weighted paths dominating optimal version is then derived as the recommended allocation for the considered instruction and deployed in the form of the recommendation bit.

\section{Evaluation}
\label{sec:eval}
\label{sec:evaluation}
In order to assess the empirical performance of the previously detailed reconfigurable register file with runtime reconfiguration, we put a main focus on evaluating the advantage of the runtime reconfiguration over a statically configured allocation mode. While runtime reconfiguration comes at overheads and the complex employment of the CFG based analysis for deriving the recommendation bits, statically configuring the register file to the horizontal or vertical allocation only could lead to a suboptimal allocation for a running application. In order to make these results well comparable, we discuss the resulting energy consumption and latency in comparison to the performance properties of SRAM.
\subsection{Setup}
In order to conduct empirical evaluation of the previously proposed register file architecture, we make full system simulations the premier evaluation target in this paper. Due to the lacking existence of market ready racetrack memory simulations, real-world simulations are infeasible. We utilize a modified version of the gem5 simulator \cite{binkert2011gem5}, which intercepts the simulation of the pipeline in order to trace out sufficient information about register accesses, which we can subsequently feed to the previously described analytic simulation model. This section briefly details the architecture of the simulation flow.

In order to derive a clean and most basic trace from gem5, we configure the simulation to run on the \emph{TimingSimple} CPU, which is a functional in-order CPU. We chose the ARM 64 bit instruction set, which sets the scope of the evaluation to a register file with $R=32$ $B=64$ bit wide registers. The internal implementation of the simple timing cpu fetches the instructions of the executed program in order. We modify the implementation of the instruction fetch, such that we parse the currently fetched instruction and write detailed information about this instruction into a trace file. We hook into the decoding process of gem5 in order to extract parsable detailed information about each single instruction. This is conducted in a two-step process: When the instruction fetch is completed, we extract the instruction address, the mnemonic and the number of source and destination registers. For each source and destination register, we trace out the register number, the accessed register width and the content of the register at the moment of the completed instruction fetch. After the execution of the fetched instruction is completed, we trace out the register content again. This allows the assembly of a single line in our trace file, containing detailed information about the executed instruction, together with register contents before and after the execution.

We further equip our simulation setup with a set of scripts to read and process trace files in the corresponding order. In order to derive the performance characteristics, we feed the accessed sequence of register into the cost model, which is also used during the offline derivation of recommendation bits. This results in a single set of performance numbers per execution of the simulation. To gain further detailed information about the quality of the derived recommendation bits by our proposed optimization method, we simulate the perofmance result of both register allocations for each block, where a new recommendation is applied. This reveals whether the correct allocation was chosen by the recommendation method and how large is the potential impact of a wrong recommendation.

\subsection{Benchmark Applications}
We aim for a diverse set of benchmark application in order to evaluate the performance of the proposed architecture. One key aspect to the performance is the amount of registers being used frequently by the target application. Only a few registers being used intensively causes different behavior in the register file than many registers being used simultaneously by the running application. Given an implementation of a program in C, the compiler controls the corresponding usage of registers. Therefore, we employ the gcc compiler with the optimization options \emph{-O0} and \emph{-O3} in the evaluation. Experimental observations reveal that \emph{-O0} tends to use fewer registers intensively while \emph{-O3} tends to use many registers simultaneously.
Furthermore, these observations reveal that a large difference exists between applying no compiler optimization and applying compiler optimization, but not between the actual optimization levels.
We implement two exemplary benchmarks in C, namely a quick sort implementation and a linear equation solver, employng Gaussian elimination. Additionally, we study a third benchmark from the area of database processing. The benchmark implements a typical workload that could occur e.g., on a microcontroller in a sensor network. In this scenario, the values of 1000 simulated sensors are updated (20 attributes with 64 bits each) and the single attributes are then aggregated. In contrast to the other two applications, we escape generation of C code and use a direct query compiler for this benchmark that synthesizes the implemented operations (update and aggregation) directly into assembly code and is therefore not subject to the optimization levels of the compiler.

Consequently, we use these versions of benchmark applications in the following:
\begin{itemize}[leftmargin=*]
    \item \emph{lesolve\_O0}: linear equation solver in C, compiled with \emph{gcc -O0}
    \item \emph{lesolve\_O3}: linear equation solver in C, compiled with \emph{gcc -O3}
    \item \emph{qsort\_O0}: quick sort in C, compiled with \emph{gcc -O0}
    \item \emph{qsort\_O3}: quick sort in C, compiled with \emph{gcc -O3}
    \item \emph{db}: database benchmark as synthesized assembly code
\end{itemize}
Each of these benchmarks leads to a register access trace, which is analyzed according to the previously described evaluation setup. In order to assess the resulting performance, the respective number of read and written registers has to be considered. These are reported in \cref{tab:numreadwrite}.
\begin{table}[h]
    \begin{tabular}{c|rrr}
        \textbf{Benchmark}&\textbf{Reads}&\textbf{Writes}&\textbf{Total}\\
        \hline
        \textbf{lesolve\_O0}&506\,465\,141&281\,100\,996&787\,566\,137\\
        \textbf{lesolve\_O3}&124\,587\,253&47\,765\,254&172\,352\,507\\
        \textbf{qsort\_O0}&9\,936\,837&7\,950\,483&17\,887\,320\\
        \textbf{qsort\_O3}&6\,258\,029&5\,112\,664&11\,370\,693\\
        \textbf{db}&147\,069\,841&86\,044\,076&233\,113\,917\\
    \end{tabular}
    \caption{Number of Register Accesses for Benchmarks}
    \vspace*{-1cm}
    \label{tab:numreadwrite}
\end{table}

\subsection{Results}
\def\shiftscaler{1}

\newcommand{\plotversion}[5]{

    \benchplot{green}{prob_../../binaries/lesolve_O0}{#1}{#2}{\shiftscaler*-36}{#3}{#4}{#5}
    \benchplot{green!60!black}{prob_../../binaries/lesolve_O3}{#1}{#2}{\shiftscaler*-18}{#3}{#4}{#5}

    \benchplot{orange}{prob_../../binaries/qsort_O0}{#1}{#2}{\shiftscaler*0}{#3}{#4}{#5}
    \benchplot{orange!80!black}{prob_../../binaries/qsort_O3}{#1}{#2}{\shiftscaler*18}{#3}{#4}{#5}
    \benchplot{purple}{prob_../../binaries/db}{#1}{#2}{\shiftscaler*36}{#3}{#4}{#5}

}
\newcommand{\helplines}[1]{}

\newcommand{\benchplot}[8]{
    \begin{scope}[shift={(#5,0)}]
        \addplot+[solid,color=#1,draw=black, fill, only marks, mark=#6,mark options={scale=#7}, y filter/.append code={
            \ifnum \conda
                \ifnum \condb
                    \ifnum \condc
                        \ifnum \condd
                            \ifnum 1=1
                                \ifthenelse{\equal{\thisrow{benchmark}}{#2}}
                                {\pgfmathparse{#3}\pgfmathresult}
                                {\def\pgfmathresult{nan}}
                            \fi
                        \else
                            \def\pgfmathresult{nan}
                        \fi
                    \else
                        \def\pgfmathresult{nan}
                    \fi
                \else
                    \def\pgfmathresult{nan}
                \fi
            \else
                \def\pgfmathresult{nan}
            \fi
        },
        ] table[col sep=comma, x=#4] {results.csv};
    \end{scope}
}

\def\plotwidth{\textwidth}

\newcommand{\compplot}[9]{
    \begin{tikzpicture}
        \begin{axis}[
            width=\plotwidth,
            height=4cm,
            ylabel=#9,
            xlabel=#8,
            ymajorgrids=true,
            yminorgrids=true,
            xmajorgrids=true,
            legend style={at={(0,-.5)},anchor=north west},
            symbolic x coords=#6,
            xtick=#7,
            ymode=log,
            log basis y=10,
        ]

            \begin{scope}[shift={(\shiftscaler*-36,0)}]
                \helplines{green!80!black}
            \end{scope}
            \begin{scope}[shift={(\shiftscaler*-18,0)}]
                \helplines{green!40!black}
            \end{scope}
            \begin{scope}[shift={(\shiftscaler*0,0)}]
                \helplines{orange!80!black}
            \end{scope}
            \begin{scope}[shift={(\shiftscaler*18,0)}]
                \helplines{orange!60!black}
            \end{scope}
            \begin{scope}[shift={(\shiftscaler*36,0)}]
                \helplines{purple!80!black}
            \end{scope}

            \plotversion{#2}{#1}{*}{1.5}{1.95}
            \plotversion{#3}{#1}{star}{1.5}{1.775}
            \plotversion{#4}{#1}{+}{1.5}{1.6}
            \plotversion{#5}{#1}{-}{1.5}{1.6}
        \end{axis}

    \end{tikzpicture}
}

\subsubsection{Starting Point: Intuitive Use Case}

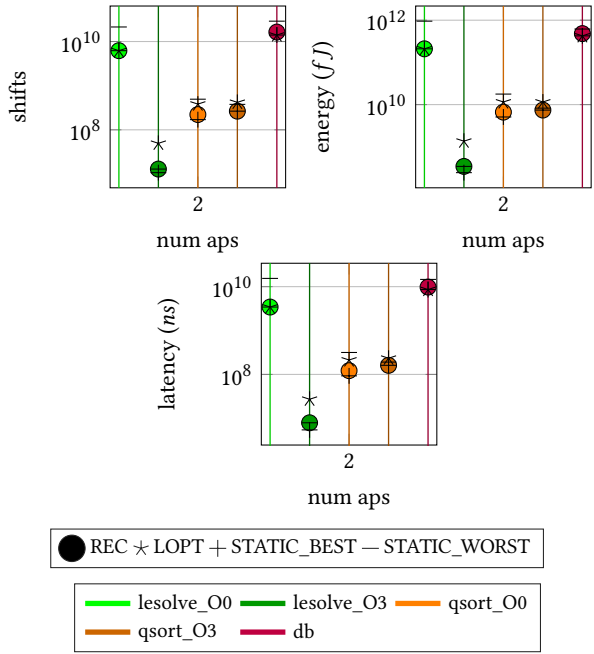
\begin{figure}[t]
    \centering

    \compplot{num_aps}
    {log10(\thisrow{recommended_total_shifts})}
    {log10(\thisrow{opt_total_shifts})}
    {log10(min(\thisrow{v1_total_shifts},\thisrow{v2_total_shifts}))}
    {log10(max(\thisrow{v1_total_shifts},\thisrow{v2_total_shifts}))}
    {{1,2,4,8,16,32,64,128,256,512,1024,2048}}
    {{2}}
    {num aps}{shifts}
    \compplot{num_aps}
    {log10(\thisrow{recommended_total_energy})}
    {log10(\thisrow{opt_total_energy})}
    {log10(min(\thisrow{v1_total_energy},\thisrow{v2_total_energy}))}
    {log10(max(\thisrow{v1_total_energy},\thisrow{v2_total_energy}))}
    {{1,2,4,8,16,32,64,128,256,512,1024,2048}}
    {{2}}
    {num aps}{energy ($fJ$)}
    \compplot{num_aps}
    {log10(\thisrow{recommended_total_latency})}
    {log10(\thisrow{opt_total_latency})}
    {log10(min(\thisrow{v1_total_latency},\thisrow{v2_total_latency}))}
    {log10(max(\thisrow{v1_total_latency},\thisrow{v2_total_latency}))}
    {{1,2,4,8,16,32,64,128,256,512,1024,2048}}
    {{2}}
    {num aps}{latency ($ns$)}

    \begin{tikzpicture} 
        \begin{axis}[%
        hide axis,
        xmin=10,
        xmax=50,
        ymin=0,
        ymax=0.4,
        legend style={draw=white!15!black,legend cell align=left},
        legend columns=4
        ]
        \addlegendimage{mark=*,only marks,mark options={scale=2.5}}
        \addlegendentry{\small REC};
        \addlegendimage{mark=star,only marks,mark options={scale=1.5}}
        \addlegendentry{\small LOPT};
        \addlegendimage{mark=+,only marks,mark options={scale=1.6}}
        \addlegendentry{\small STATIC\_BEST};
        \addlegendimage{mark=-,only marks,mark options={scale=1.6}}
        \addlegendentry{\small STATIC\_WORST};
        \end{axis}
    \end{tikzpicture}

    \vspace*{-5cm}

    \begin{tikzpicture} 
        \begin{axis}[%
        hide axis,
        xmin=10,
        xmax=50,
        ymin=0,
        ymax=0.4,
        legend style={draw=white!15!black,legend cell align=left},
        legend columns=3
        ]
        \addlegendimage{mark=none,color=green,line width=2}
        \addlegendentry{\small lesolve\_O0};
        \addlegendimage{mark=none,color=green!60!black,line width=2}
        \addlegendentry{\small lesolve\_O3};
        \addlegendimage{mark=none,color=orange,line width=2}
        \addlegendentry{\small qsort\_O0};
        \addlegendimage{mark=none,color=orange!80!black,line width=2}
        \addlegendentry{\small qsort\_O3};
        \addlegendimage{mark=none,color=purple,line width=2}
        \addlegendentry{\small db};
        \end{axis}
    \end{tikzpicture}
    \vspace*{-5cm}

    \caption{Shift, Energy and Latency of an Intuitive Configuration}
    \label{fig:intuitive}
\end{figure}

The first configuration to be addressed in this evaluation is an intuitive register file assembly where all parameters are fixed. Towards this, we make an intuitive allocation of one register to one nanotrack. This leads to 64 position wide nanotracks with 32 nanotracks in total. We equip the register file with two access ports in order to make one register fully acccesible at one aligned position in the vertical allocation. The window size for the allocation recommendations is set to 100 instructions:\\
\begin{tikzpicture}
    \node[draw=green!80!black, line width=2, minimum width=.47\textwidth] () at (0,0) {
        \begin{minipage}{.47\textwidth}
            \centering
            \begin{tabular}{c|c|c|c}
                \textbf{Width}&\textbf{\# Tracks}&\textbf{\# APs}&\textbf{Window}\\
                \hline
                64&32&2&100
            \end{tabular}
        \end{minipage}
    };
\end{tikzpicture}
\\
We evaluate the required number of shifts, the total energy consumption and the total spent latency for all accesses for each benchmark configuration. Along that, we consider 4 versions of runtime register file configurations to compare: 1) the deployed recommendations being applied (\textbf{REC}), 2) the local optimal (based on ideal a priori knowledge) allocation to be applied for every window (\textbf{LOPT}), 3) statically configuring the register file to the best of the horizontal of vertical allocation (based on ideal a priori knowledge) during the entire benchmark (\textbf{STATIC\_BEST}) and 4) statically configuring the register file to the worst of the horizontal or vertical allocation (\textbf{STATIC\_WORST}). The three baseline version (\textbf{OPT},\textbf{STATIC\_BEST} and \textbf{STATIC\_WORST}) are not realizable due to the required a priori knowledge and can only be conducted in the simulation. However, they reveal a relative assessment of the derived recommendations. \cref{fig:intuitive} depicts the corresponding results.

\def\conda{\thisrow{window_size}=100}
\def\condb{\thisrow{num_tracks}=32}
\def\condc{\thisrow{track_length}=64}
\def\condd{\thisrow{num_aps}=2}
\def\plotwidth{.22\textwidth}
\def\shiftscaler{0.5}
\renewcommand{\helplines}[1]{
    \draw[#1, line width=.5](axis cs: 2,-1)--(axis cs: 2,10E19);*
}
\def\plotwidth{\textwidth}
\def\shiftscaler{1}

In the depicted figure, the required amount of shifts, the allover energy consumption and the total latency can be seen from left to the right. Within each of the subplots, each benchmark application is displayed at one y-axis parallel line. Along the line, the baseline versions and the recommendation version are indicated by different markers. First, it can be observed that the recommendation version never underperforms the static worst allocation. This means, in case the register file would be operated with a static allocation only, it can be worse than the recommendation version. Further, it can be seen that the local optimal version performs worse in terms of shifts for some cases. This can be caused by the fact that due to the local optimization also overheads for reconfiguring the allocation are introduced. In the recommendation version, these overheads are introduced as well, but considered during the CFG based optimization. Generally, it can be seen that choosing a static allocation only could lead to high performance in the best case but also to poor performance in the worst case. This suggests the conclusion that runtime reconfiguration are crucially important. Given this, the recommendation version can even outperform the local optimal version for several cases.

Considering the results beyond a relative comparison, the average shift, energy and latency per register access has to be computed. We compare the static worst allocation (\cref{tab:avgintsw}) to the allocation derived by the recommendations (\cref{tab:avgintrec}). The static worst allocation indicates the possible bad case when a static allocation is chosen instead of a dynamic approach.
\begin{table}[h]
    \begin{tabular}{c|rrr}
        \textbf{Benchmark}&\textbf{avg. Shift}&\textbf{avg. Energy}&\textbf{avg. Latency}\\
        \hline
        \textbf{lesolve\_O0}&
        7.88
        &
        266.21 $fJ$
        &
        4.35 $nS$
        \\
        \textbf{lesolve\_O3}&
        0.07
        &
        1.99 $fJ$
        &
        0.05 $nS$
        \\
        \textbf{qsort\_O0}&
        10.04
        &
        299.96 $fJ$
        &
        5.48 $nS$
        \\
        \textbf{qsort\_O3}&
        23.28
        &
        652.40 $fJ$
        &
        14.16 $nS$
        \\
        \textbf{db}&
        60.13
        &
        1824.30 $fJ$
        &
        36.86 $nS$
        \\
    \end{tabular}
    \caption{Average Energy and Latency for an Intuitive Configuration: REC}
    \label{tab:avgintrec}
\end{table}
\begin{table}[h]
    \begin{tabular}{c|rrr}
        \textbf{Benchmark}&\textbf{avg. Shift}&\textbf{avg. Energy}&\textbf{avg. Latency}\\
        \hline
        \textbf{lesolve\_O0}&
        27.11
        &
        1204.10 $fJ$
        &
        19.57 $nS$
        \\
        \textbf{lesolve\_O3}&
        0.07
        &
        1.99 $fJ$
        &
        0.05 $nS$
        \\
        \textbf{qsort\_O0}&
        27.62
        &
        992.75 $fJ$
        &
        17.68 $nS$
        \\
        \textbf{qsort\_O3}&%
        33.16
        &
        725.13 $fJ$
        &
        16.79 $nS$
        \\
        \textbf{db}&%
        124.00
        &
        2647.38 $fJ$
        &
        62.70 $nS$
        \\
    \end{tabular}
    \caption{Average Energy and Latency for an Intuitive Configuration: STATIC\_WORST}
    \label{tab:avgintsw}
\end{table}
In order to assess these number further, we consider corresponding average latency and energy consumption of SRAM, as reported in \cite{oh2016power}. Due to the similarity in read and write procedures in SRAM, we account the same range in terms of latency for both operations. The numbers are depicted in \cref{tab:sramdata}.
\begin{table}[h]
    \begin{tabular}{c|rr}
        \textbf{Operation}&\textbf{Energy / Access}&\textbf{Latency / Access}\\
        \hline
        \textbf{Read}&$0.39-0.71 pJ$&$164-254 ns$\\
        \textbf{Write}&$0.8-1.57 pJ$&$164-254 ns$\\
    \end{tabular}
    \caption{Average Latency and Energy Consumption for SRAM \cite{oh2016power}}
    \label{tab:sramdata}
\end{table}
While even under the static worst case the racetrack register file outperforms SRAM by far in terms of latency, the energy consumption allows for interesting observations. When a static approach is chosen, the worst case can easily lead to energy consumption worse than SRAM. When the recommendation based dynamic approach is chosen, the energy consumption can be reduced even below SRAM in most cases.
This suggests the conclusion that in addition to superior latency performance of our proposed racetrack register file, dynamic allocation delivers a crucial component to make the register file highly competitive in terms of the energy consumption.

\subsubsection{Exploration: Number of Access Ports}

Going beyond the intuitive configuration, picked up as a starting point before, we explore the configuration degrees of the register file architecture in the following. Therefore, we keep the majority of parameters as before and vary one of them. This is being done with the number of access ports in the following. Consequently, the nanotrack width remains at 64 positions, the number of nanotracks remains at 32 and the recommendation window size remains at 100:
\\
\begin{tikzpicture}
    \node[draw=green!80!black, line width=2, minimum width=.47\textwidth] () at (0,0) {
        \begin{minipage}{.47\textwidth}
            \centering
            \begin{tabular}{c|c|c|c}
                \textbf{Width}&\textbf{\# Tracks}&\textbf{\# APs}&\textbf{Window}\\
                \hline
                64&32&2-64&100
            \end{tabular}
        \end{minipage}
    };
\end{tikzpicture}
\\
The same baseline comparisons are used as before. We further utilize the same plot format as before and denote the number of access ports along the x-axis.
\def\conda{\thisrow{window_size}=100}
\def\condb{\thisrow{num_tracks}=32}
\def\condc{\thisrow{track_length}=64}
\def\condd{1=1}
\renewcommand{\helplines}[1]{
    \draw[#1, line width=.5](axis cs: 2,-1)--(axis cs: 2,10E19);
    \draw[#1, line width=.5](axis cs: 4,-1)--(axis cs: 4,10E19);
    \draw[#1, line width=.5](axis cs: 8,-1)--(axis cs: 8,10E19);
    \draw[#1, line width=.5](axis cs: 16,-1)--(axis cs: 16,10E19);
    \draw[#1, line width=.5](axis cs: 32,-1)--(axis cs: 32,10E19);
    \draw[#1, line width=.5](axis cs: 64,-1)--(axis cs: 64,10E19);
}
\def\plotwidth{.47\textwidth}
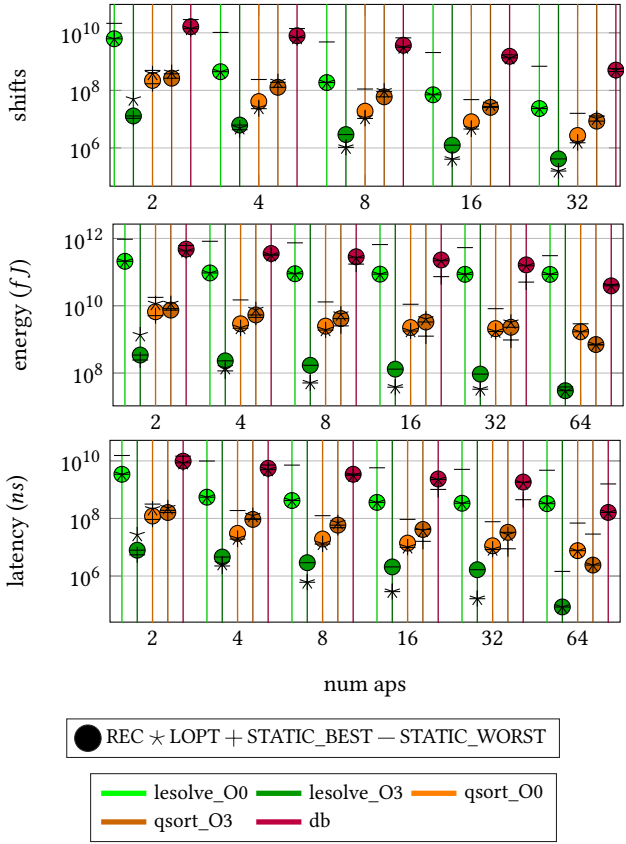
\begin{figure}
    \centering

    \compplot{num_aps}
    {log10(\thisrow{recommended_total_shifts})}
    {log10(\thisrow{opt_total_shifts})}
    {log10(min(\thisrow{v1_total_shifts},\thisrow{v2_total_shifts}))}
    {log10(max(\thisrow{v1_total_shifts},\thisrow{v2_total_shifts}))}
    {{1,2,4,8,16,32,64,128,256,512,1024,2048}}
    {{2,4,8,16,32,64}}
    {}
    {shifts}
    \compplot{num_aps}
    {log10(\thisrow{recommended_total_energy})}
    {log10(\thisrow{opt_total_energy})}
    {log10(min(\thisrow{v1_total_energy},\thisrow{v2_total_energy}))}
    {log10(max(\thisrow{v1_total_energy},\thisrow{v2_total_energy}))}
    {{1,2,4,8,16,32,64,128,256,512,1024,2048}}
    {{2,4,8,16,32,64}}
    {}
    {energy ($fJ$)}
    \compplot{num_aps}
    {log10(\thisrow{recommended_total_latency})}
    {log10(\thisrow{opt_total_latency})}
    {log10(min(\thisrow{v1_total_latency},\thisrow{v2_total_latency}))}
    {log10(max(\thisrow{v1_total_latency},\thisrow{v2_total_latency}))}
    {{1,2,4,8,16,32,64,128,256,512,1024,2048}}
    {{2,4,8,16,32,64}}
    {\strut num aps}{latency ($ns$)}

    \begin{tikzpicture} 
        \begin{axis}[%
        hide axis,
        xmin=10,
        xmax=50,
        ymin=0,
        ymax=0.4,
        legend style={draw=white!15!black,legend cell align=left},
        legend columns=4
        ]
        \addlegendimage{mark=*,only marks,mark options={scale=2.5}}
        \addlegendentry{\small REC};
        \addlegendimage{mark=star,only marks,mark options={scale=1.5}}
        \addlegendentry{\small LOPT};
        \addlegendimage{mark=+,only marks,mark options={scale=1.6}}
        \addlegendentry{\small STATIC\_BEST};
        \addlegendimage{mark=-,only marks,mark options={scale=1.6}}
        \addlegendentry{\small STATIC\_WORST};
        \end{axis}
    \end{tikzpicture}

    \vspace*{-5cm}

    \begin{tikzpicture} 
        \begin{axis}[%
        hide axis,
        xmin=10,
        xmax=50,
        ymin=0,
        ymax=0.4,
        legend style={draw=white!15!black,legend cell align=left},
        legend columns=3
        ]
        \addlegendimage{mark=none,color=green,line width=2}
        \addlegendentry{\small lesolve\_O0};
        \addlegendimage{mark=none,color=green!60!black,line width=2}
        \addlegendentry{\small lesolve\_O3};
        \addlegendimage{mark=none,color=orange,line width=2}
        \addlegendentry{\small qsort\_O0};
        \addlegendimage{mark=none,color=orange!80!black,line width=2}
        \addlegendentry{\small qsort\_O3};
        \addlegendimage{mark=none,color=purple,line width=2}
        \addlegendentry{\small db};
        \end{axis}
    \end{tikzpicture}
    \vspace*{-5cm}

\caption{Varying Number of Access Ports}
\label{fig:varnumap}
\end{figure}
\cref{fig:varnumap} depicts the corresponding result data for the number of access varying from 1 to 32. Overall, a similar observation as before can be made, namely that the worst static allocation leads to the highest amount of shifts, energy consumption and latency in most cases. Although the best static allocation, as well as the locally optimal allocation outperforms the recommended version in several cases and is close the recommended version in many cases, the conclusion is suggested that the recommendation based allocation can deliver a good allocation in most cases, which optimizes beyond the worst static allocation. Over the increasing number of access ports, the number of shifts, the energy consumption and the latency are continuously reduced. Picking the example of lesolve\_O0, shifts reduce from 2 to 32 access ports by 
$264\times$ down to $0.03$ per access, energy consumption by 
$2.4\times$ down to $109.25 fJ$ per access and latency by 
$10.1\times$ down to $0.43 nS$ per access for the recommendation based allocation, respectively.
The static worst allocation at the same time is reduced by 
$31\times$ down to $0.87$ shifts per access, energy consumption is reduced by 
$1.77\times$ down to $679.33 fJ$ per access and latency is reduced by 
$3.03\times$ down to $6.45 nS$ per access, respectively.
This further supports the advantage of the dynamic recommendation based allocation approach. While the energy consumption of the static worst allocation still hardly competes with SRAM, the dynamic recommendation based version strongly outperforms SRAM in terms of energy consumption.
Although these results clearly suggest an increase in the number of access ports to the maximal level to be clearly favorable in terms of shifts, energy and latency, it has to be noted that employing additional access ports comes at the cost of additional and more complex hardware components. However, the maximally feasible amount should be considered.

\subsubsection{Exploration: Window Size}

After varying the number of access ports, the next interesting parameter is the window size for the recommendation mechanism. A smaller window size can lead to a more accurate runtime configuration on the cost of a potentially larger overhead for switching between allocations. Hence, we keep the width of the nanotracks at 64, the number of nanotracks at 32, the number of access ports at 8 and vary the window size in the following:
\\
\begin{tikzpicture}
    \node[draw=green!80!black, line width=2, minimum width=.47\textwidth] () at (0,0) {
        \begin{minipage}{.47\textwidth}
            \centering
            \begin{tabular}{c|c|c|c}
                \textbf{Width}&\textbf{\# Tracks}&\textbf{\# APs}&\textbf{Window}\\
                \hline
                64&32&8&10-2000
            \end{tabular}
        \end{minipage}
    };
\end{tikzpicture}
\\
The plot format is kept as before with the varying window size being depicted along the x-axis.
\def\conda{\thisrow{num_aps}=8}
\def\condb{\thisrow{num_tracks}=32}
\def\condc{\thisrow{track_length}=64}
\def\condd{1=1}
\def\shiftscaler{0.6}
\renewcommand{\helplines}[1]{
    \draw[#1, line width=.5](axis cs: 10,-1)--(axis cs: 10,10E19);
    \draw[#1, line width=.5](axis cs: 100,-1)--(axis cs: 100,10E19);
    \draw[#1, line width=.5](axis cs: 1000,-1)--(axis cs: 1000,10E19);
    \draw[#1, line width=.5](axis cs: 2000,-1)--(axis cs: 2000,10E19);
}
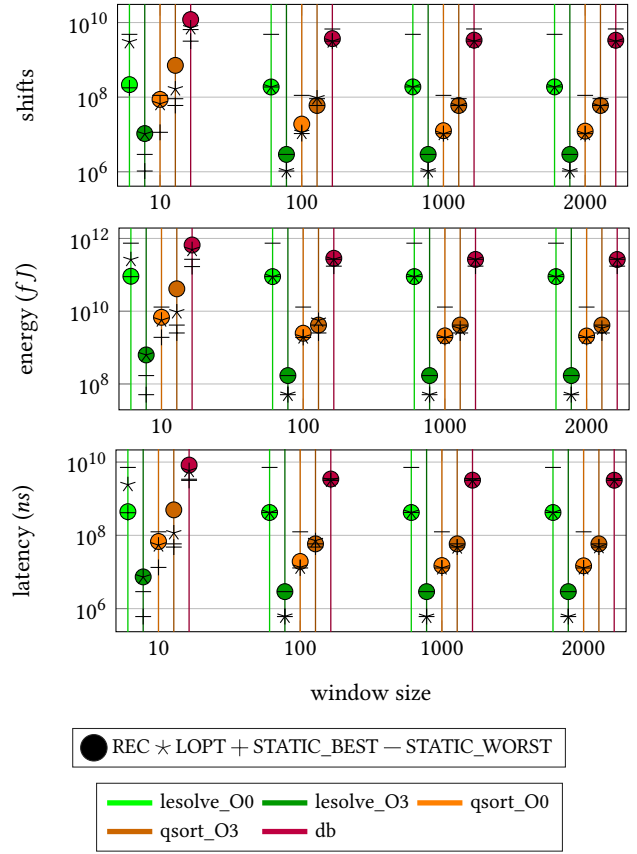
\begin{figure}
    \centering

    \compplot{window_size}
    {log10(\thisrow{recommended_total_shifts})}
    {log10(\thisrow{opt_total_shifts})}
    {log10(min(\thisrow{v1_total_shifts},\thisrow{v2_total_shifts}))}
    {log10(max(\thisrow{v1_total_shifts},\thisrow{v2_total_shifts}))}
    {{10,100,1000,2000}}
    {{10,100,1000,2000}}
    {}
    {shifts}
    \compplot{window_size}
    {log10(\thisrow{recommended_total_energy})}
    {log10(\thisrow{opt_total_energy})}
    {log10(min(\thisrow{v1_total_energy},\thisrow{v2_total_energy}))}
    {log10(max(\thisrow{v1_total_energy},\thisrow{v2_total_energy}))}
    {{10,100,1000,2000}}
    {{10,100,1000,2000}}
    {}
    {energy ($fJ$)}
    \compplot{window_size}
    {log10(\thisrow{recommended_total_latency})}
    {log10(\thisrow{opt_total_latency})}
    {log10(min(\thisrow{v1_total_latency},\thisrow{v2_total_latency}))}
    {log10(max(\thisrow{v1_total_latency},\thisrow{v2_total_latency}))}
    {{10,100,1000,2000}}
    {{10,100,1000,2000}}
    {\strut window size}{latency ($ns$)}

    \begin{tikzpicture} 
        \begin{axis}[%
        hide axis,
        xmin=10,
        xmax=50,
        ymin=0,
        ymax=0.4,
        legend style={draw=white!15!black,legend cell align=left},
        legend columns=4
        ]
        \addlegendimage{mark=*,only marks,mark options={scale=2.5}}
        \addlegendentry{\small REC};
        \addlegendimage{mark=star,only marks,mark options={scale=1.5}}
        \addlegendentry{\small LOPT};
        \addlegendimage{mark=+,only marks,mark options={scale=1.6}}
        \addlegendentry{\small STATIC\_BEST};
        \addlegendimage{mark=-,only marks,mark options={scale=1.6}}
        \addlegendentry{\small STATIC\_WORST};
        \end{axis}
    \end{tikzpicture}

    \vspace*{-5cm}

    \begin{tikzpicture} 
        \begin{axis}[%
        hide axis,
        xmin=10,
        xmax=50,
        ymin=0,
        ymax=0.4,
        legend style={draw=white!15!black,legend cell align=left},
        legend columns=3
        ]
        \addlegendimage{mark=none,color=green,line width=2}
        \addlegendentry{\small lesolve\_O0};
        \addlegendimage{mark=none,color=green!60!black,line width=2}
        \addlegendentry{\small lesolve\_O3};
        \addlegendimage{mark=none,color=orange,line width=2}
        \addlegendentry{\small qsort\_O0};
        \addlegendimage{mark=none,color=orange!80!black,line width=2}
        \addlegendentry{\small qsort\_O3};
        \addlegendimage{mark=none,color=purple,line width=2}
        \addlegendentry{\small db};
        \end{axis}
    \end{tikzpicture}
    \vspace*{-5cm}

\caption{Varying Window Size}
\label{fig:varws}
\end{figure}
\cref{fig:varws} depicts the corresponding results. It can be observed that the data points for all window sizes above $10$ are extremely similar. This suggests the conclusion that an extension of the window likely only covers repetitive execution of the same instruction sequence and therefore makes no difference to neither the recommendation, nor to the introduced overheads.
For the window size of 10 instructions, it can be observed that the local optimal version, as well as the recommendation version cause an increased number of shifts, energy consumption and latency. This can be explained by the additional overhead of potentially changing the allocation on a granularity of 10 instructions. This is further supported by the fact that for a window size of 10 the static worst and best allocation outperforms the local optimal and recommendation version by far.

\subsubsection{Exploration: Number of Nanotracks} \label{sec:evaluation-nanotracks}

A last degree of variation left is the number of nanotracks, which is tightly coupled with their width. As the register file always has to serve the capacity of 32 registers with 64 bits each, a total of 2048 bits need to be provided. Given a certain number of nanotracks, this defines their width. Consequently, we vary the number of nanotracks between 8 and 256, which consequently requires nanotrack widths of 256-8. We keep the number of access ports fixed at 8 and the window size for recommendations at 100:
\\
\begin{tikzpicture}
    \node[draw=green!80!black, line width=2, minimum width=.47\textwidth] () at (0,0) {
        \begin{minipage}{.47\textwidth}
            \centering
            \begin{tabular}{c|c|c|c}
                \textbf{Width}&\textbf{\# Tracks}&\textbf{\# APs}&\textbf{Window}\\
                \hline
                8-256&8-256&8&100
            \end{tabular}
        \end{minipage}
    };
\end{tikzpicture}
\\
\def\conda{\thisrow{num_aps}=8}
\def\condb{\thisrow{window_size}=100}
\def\condc{1=1}
\def\condd{1=1}
\def\shiftscaler{1}
The plot format is again kept as before. The number of nanotracks is varied along the x-axis. The corresponding width of the nanotracks is not explicitly shown in the figure, but fixed due to the aforementioned constraint.
\renewcommand{\helplines}[1]{
    \draw[#1, line width=.5](axis cs: 8,-1)--(axis cs: 8,10E19);
    \draw[#1, line width=.5](axis cs: 16,-1)--(axis cs: 16,10E19);
    \draw[#1, line width=.5](axis cs: 32,-1)--(axis cs: 32,10E19);
    \draw[#1, line width=.5](axis cs: 64,-1)--(axis cs: 64,10E19);
    \draw[#1, line width=.5](axis cs: 128,-1)--(axis cs: 128,10E19);
    \draw[#1, line width=.5](axis cs: 256,-1)--(axis cs: 256,10E19);
}
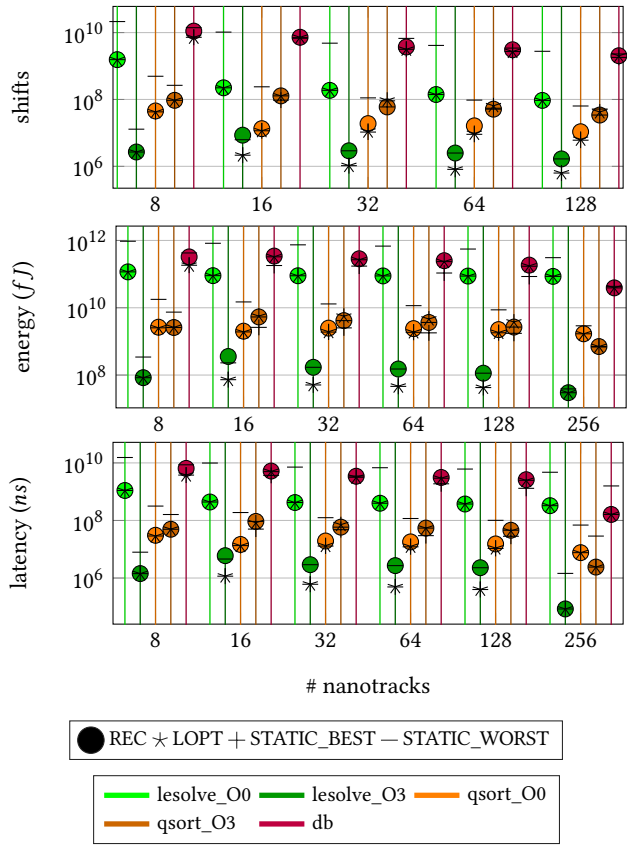
\begin{figure}
    \centering

    \compplot{num_tracks}
    {log10(\thisrow{recommended_total_shifts})}
    {log10(\thisrow{opt_total_shifts})}
    {log10(min(\thisrow{v1_total_shifts},\thisrow{v2_total_shifts}))}
    {log10(max(\thisrow{v1_total_shifts},\thisrow{v2_total_shifts}))}
    {{1,2,4,8,16,32,64,128,256,512,1024,2048}}
    {{1,2,4,8,16,32,64,128,256,512,1024,2048}}
    {}
    {shifts}
    \compplot{num_tracks}
    {log10(\thisrow{recommended_total_energy})}
    {log10(\thisrow{opt_total_energy})}
    {log10(min(\thisrow{v1_total_energy},\thisrow{v2_total_energy}))}
    {log10(max(\thisrow{v1_total_energy},\thisrow{v2_total_energy}))}
    {{1,2,4,8,16,32,64,128,256,512,1024,2048}}
    {{1,2,4,8,16,32,64,128,256,512,1024,2048}}
    {}
    {energy ($fJ$)}
    \compplot{num_tracks}
    {log10(\thisrow{recommended_total_latency})}
    {log10(\thisrow{opt_total_latency})}
    {log10(min(\thisrow{v1_total_latency},\thisrow{v2_total_latency}))}
    {log10(max(\thisrow{v1_total_latency},\thisrow{v2_total_latency}))}
    {{1,2,4,8,16,32,64,128,256,512,1024,2048}}
    {{1,2,4,8,16,32,64,128,256,512,1024,2048}}
    {\strut \# nanotracks}{latency ($ns$)}

    \begin{tikzpicture} 
        \begin{axis}[%
        hide axis,
        xmin=10,
        xmax=50,
        ymin=0,
        ymax=0.4,
        legend style={draw=white!15!black,legend cell align=left},
        legend columns=4
        ]
        \addlegendimage{mark=*,only marks,mark options={scale=2.5}}
        \addlegendentry{\small REC};
        \addlegendimage{mark=star,only marks,mark options={scale=1.5}}
        \addlegendentry{\small LOPT};
        \addlegendimage{mark=+,only marks,mark options={scale=1.6}}
        \addlegendentry{\small STATIC\_BEST};
        \addlegendimage{mark=-,only marks,mark options={scale=1.6}}
        \addlegendentry{\small STATIC\_WORST};
        \end{axis}
    \end{tikzpicture}

    \vspace*{-5cm}

    \begin{tikzpicture} 
        \begin{axis}[%
        hide axis,
        xmin=10,
        xmax=50,
        ymin=0,
        ymax=0.4,
        legend style={draw=white!15!black,legend cell align=left},
        legend columns=3
        ]
        \addlegendimage{mark=none,color=green,line width=2}
        \addlegendentry{\small lesolve\_O0};
        \addlegendimage{mark=none,color=green!60!black,line width=2}
        \addlegendentry{\small lesolve\_O3};
        \addlegendimage{mark=none,color=orange,line width=2}
        \addlegendentry{\small qsort\_O0};
        \addlegendimage{mark=none,color=orange!80!black,line width=2}
        \addlegendentry{\small qsort\_O3};
        \addlegendimage{mark=none,color=purple,line width=2}
        \addlegendentry{\small db};
        \end{axis}
    \end{tikzpicture}
    \vspace*{-5cm}

\caption{Varying Number of Nanotracks}
\label{fig:varnts}
\end{figure}
The corresponding results can be seen in \cref{fig:varnts}. Similar observations as for the increasing number of access ports can be made. With an increasing number of nanotracks, the amount of shifts, the energy consumption and the latency is continuously decreased. 
This stems from the fact that the width of the nanotracks decreases together with their increased number. This makes more registers available at the access ports in parallel and therefore causes fewer overheads.
Also, similar to the previous results, the recommendation version and the local optimal version outperform the worst static allocation, while underperforming the best static allocation in most cases. This supports the previously made conclusions.
Picking the example of lesolve\_O0 again, shifts reduce from between 8 and 128 nanotracks by 
$16.56\times$ down to $0.12$ per access, energy consumption by 
$1.33\times$ down to $111.03 fJ$ per access and latency by 
$2.95\times$ down to $0.47 nS$ per access, respectively.
The results generally suggest the conclusion that in terms of induced shifts, as many nanotracks should be chosen as possible. The improvement in terms of energy consumption and latency, however, is not as strong as for the shifts. This stems from the higher overhead due to the larger access ports, that have to span across all nanotracks.
Consequently, additional hardware costs due to the additional nanotracks should be carefully traded of against the potential gains in terms of energy consumption and latency by the increase number of nanotracks.

\section{Conclusion}
In this paper, we propose an architecture for a CPU register file, consisting of racetrack memory. The inherent properties of racetrack memories allow a trade-off for the actual allocation of registers to physical memory domains. We propose an architecture, which allows a runtime reconfiguration of this allocation. The thereby provided optimization problem of choosing the best runtime allocation is tackled by conducting probability enhanced static analysis on the target applications, featuring a detailed cost model for the proposed register file. Experimental evaluation shows that runtime reconfiguration is crucially needed since the performance of the proposed register file can degrade below SRAM in case no reconfiguration is done. Yet,  we show that it can outperform SRAM by far in case of properly done runtime reconfiguration, as provided by our application analysis.

\clearpage
\bibliographystyle{ACM-Reference-Format}
\bibliography{references}

\end{document}